\documentclass[aps,prb,reprint,groupedaddress]{revtex4-2}

\usepackage{amsmath}
\usepackage{amssymb}
\usepackage{mathrsfs}

\usepackage{txfonts}
\usepackage{graphicx}

\begin{document}

\title{Frustrations on decorated planar lattices in Ising model}

\author{F.A. Kassan-Ogly}
\email{Felix.Kassan-Ogly@imp.uran.ru}
\affiliation{M.N. Mikheev Institute of Metal Physics of Ural Branch of Russian Academy of Sciences, S. Kovalevskoy Street 18, 620108 Ekaterinburg, Russia}

\author{A.V. Zarubin}
\email{Alexander.Zarubin@imp.uran.ru}
\affiliation{M.N. Mikheev Institute of Metal Physics of Ural Branch of Russian Academy of Sciences, S. Kovalevskoy Street 18, 620108 Ekaterinburg, Russia}


\begin{abstract}
We study the frustration properties of the Ising model on several decorated lattices with arbitrary numbers of decorating spins on all bonds of the lattice within an exact analytical approach based on the Kramers--Wannier transfer-matrix technique. The existence of magnetic frustrations in such situations and their influence on the behavior of the thermodynamic functions of systems is shown. The most important result of our study is related to the description of the possible coexistence of frustrations and long-range magnetic order in partially ordered spin systems.
\end{abstract}

\maketitle

\section{Introduction}

Frustrated spin systems are being investigated nowadays exceedingly intensively. A great deal of research interest has resulted in a rather impressive list of existing literature that continues to replenish every year~\cite{Kassan-Ogly:2010:,Balents:2010,Lacroix:2011,Kudasov:2012:,Diep:2013,Vasiliev:2018:,Markina:2021:}. In spite of the fact that the concept of frustration was introduced by Toulouse only in 1977~\cite{Toulouse:1977:1}, the frustrations first revealed themselves in the Ising model on the antiferromagnetic triangular lattice by Wannier in 1950~\cite{Wannier:1950}, and on the antiferromagnetic kagome lattice by Kan\^{o} and Naya in 1953~\cite{Kano:1953} by complete suppression of the phase transition.

Since then and up to now frustration phenomena have been studied not only in the Ising model, but in several other basic models of magnetism, namely, in the Potts model \cite{Matsuda:1982,Qin:2014,Kotecky:2014,Farnell:2018}, Heisenberg model \cite{Suttner:2014,Hirose:2018,Natori:2019,Huang:2021}, XXZ model \cite{Yao:2008}, and also in Hubbard model~\cite{Nourse:2021,Batista:2003}.

Frustration properties of spin systems found in theoretical works and numerical experiments, originated from the 2D triangular and kagome lattices, attracted a riveted attention to the real 3D materials with such layers. These materials, do not order down to the lowest measured temperature despite the existence of a large exchange interaction. In the paper by Coldea et~al.~\cite{Coldea:2003} the isotropic 2D triangular frustrated quantum magnet Cs$_{2}$CuCl$_{4}$ is discussed as a candidate for a quantum spin liquid. In the paper by Shimizu et~al.~\cite{Shimizu:2003} the organic insulator $\kappa $-(ET)$_{2}$Cu$_{2}$CN$_{3}$ with a nearly isotropic triangular lattice is studied, and the obtained results show that a quantum spin liquid state is likely realized in the neighborhood of the superconducting phase. The most surge of researchers interest is attracted to archetypal material herbertsmithite ZnCu$_{3}$(OH)$_{6}$Cl$_{2}$ and its close relatives: botallackite, atacamite, clinoatacamite, claringbullite, carlowite, bobkingite, tondiite, kapellasite, haydeeite, and Zn-brochantite, each of which has layered structure with Cu magnetic ions, forming an ideal kagome lattice without ordering or spin freezing behavior down to the lowest measured temperature (see, for example, the paper by Huang et~al.~\cite{Huang:2021} and the review by Norman~\cite{Norman:2016} with bibliography list of 133~items).

Obviously, in the study of frustrated systems, established theories, numerical simulation methods as well as experimental techniques have encountered a number of difficulties. In particular, these problems are tightly related to the appearance of nonzero value of the entropy in the ground state (residual entropy) in a frustrated system. However, contrary to the heat capacity, which is the physical quantity that can be measured in experiment the entropy is by no means the observable, but rather computable quantity. Usually in experiment, the entropy computation is realized by integrating the heat capacity with subsequent choice of integration constant as often as being zero. This is true in the majority of cases since many systems are not frustrated. The same is done in numerical simulation experiments. In frustrated systems, the situation is more subtle. The integration constant can not be chosen, it should be calculated. However, this is possible only in very restricted cases, either in 1D lattices in a number of models or in 2D lattices \emph{only} in the Ising model, where the exact solution to the Helmholtz free energy is obtained. This problem was widely discussed in book~\cite{Diep:2013}, as well as in the proceedings of the biennial International Conference on Highly Frustrated Magnetism (China, Shanghai, August 2021).

In turn, the concept of a decorated lattice, related to the magnetic Ising model, was originally proposed in 1951 by Sy\^{o}zi~\cite{Syozi:1951}. It consists in introducing an extra (decorating) spin on every bond between the sites of the original lattice (nodal spins). Actually Sy\^{o}zi introduced a very new transformation (the so-called decoration--iteration transformation) by means of which the thermodynamic properties of the decorated lattice can be deduced from those of the original undecorated one. The decoration--iteration transformation was generalized to multiple decorations with arbitrary number of extra spins on every bond of the original lattice~\cite{Miyazima:1968,Syozi:1972}.

The term \emph{decoration}, which merely means an insertion of extra spins, and although widely used in the low-dimensional physics, can easily be applied to real 3D crystals. For example, the bcc and fcc structures may be called the body-decorated and face-decorated simple cubic lattice. Moreover, the researchers in the solid state physics often use the term \emph{intercalation}, which also means an insertion of extra atoms to ameliorate initial material, or to obtain a material with new properties. Thus \emph{decoration} and \emph{intercalation} are simply the synonyms, and we have the terminological question at issue.

There are also a sufficient number of publications in which exact solutions were obtained and a theoretical analysis of models was carried out for studying real decorated spin systems, for example, such as 
one-dimensional mixed-spin chains CuNi(EDTA)$\cdot$6H$_{2}$O~\cite{Strecka:2011};
polymeric coordination compounds (bond-alternating chain) Cu(3-Chloropyridine)$_{2}$(N$_{3}$)$_{2}$~\cite{Strecka:2005,Torrico:2018};
diamond chain materials Cu$_{3}$(CO$_{3}$)$_{2}$(OH)$_{2}$, and Cu$_{3}$(TeO$_{3}$)$_{2}$Br$_{2}$~\cite{Canova:2009};
quasi-one-dimensional Haldane compounds $R_{2}$BaNiO$_{5}$ ($R{}={}$Y$^{3+}$, Nd$^{3+}$, Pr$^{3+}$)~\cite{Takushima:2000};
layered mixed-spin rare-earth nickelates $R_{2}$BaNiO$_{3}$ ($R$ is a magnetic rare-earth ion)~\cite{Oitmaa:2005};
two-dimensional magnetic materials Cu$_{9}X_{2}$(cpa)$_{6}\cdot x$H$_{2}$O (cpa${}={}$2-carboxypentonic acid; $X{}={}$F, Cl, Br)~\cite{Rojas:2009};
layered compound SrCo$_{6}$O$_{11}$~\cite{Galisova:2011};
two-dimensional decorated ferrimagnetic systems~\cite{Dakhama:1998,Jascur:1998}.

To study the thermodynamic and magnetic characteristics of decorated frustrated spin systems, we derived the principal eigenvalues in canonic Kramers--Wannier transfer-matrix form for decorated square, triangular, and honeycomb Ising lattices and also the spontaneous magnetizations generalized by decorations. The detailed studies were confined by isotropic cases, when the numbers of decorating spins and exchange interactions are equal in all lattice directions.

\section{Thermodynamic functions of the models}

We considered the thermodynamic properties of the generalized Ising model of a two-dimensional spin system on square, triangular, and honeycomb decorated lattices. 

The Hamiltonians of such a system for a square decorated lattice have the form 
\begin{multline}
\mathscr{H}=-\sum_{i=1}^{N_{1}}\sum_{j=1}^{N_{2}}(J_{1}\sigma_{i,j}\sigma_{i+1,j}+J_{2}\sigma_{i,j}\sigma_{i,j+1})\\
-\sum_{i=1}^{N_{1}}\sum_{j=1}^{N_{2}}\left(\sum_{m=0}^{d_{1}}J_{1}^{\prime}\sigma_{i+m,y}\sigma_{i+m+1,j}+\sum_{j=0}^{d_{2}}J_{2}^{\prime}\sigma_{i,j+n}\sigma_{i,j+n+1}\right),\label{eq:H:T3D}
\end{multline}
and for triangular and honeycomb decorated lattices 
\begin{multline}
\mathscr{H}=-\sum_{i=1}^{N_{1}}J_{1}\sigma_{i}\sigma_{i+1}-\sum_{j=0}^{d_{1}}J_{1}^{\prime}\sigma_{j}\sigma_{j+1}\\
-\sum_{k=1}^{N_{2}}J_{2}\sigma_{k}\sigma_{k+1}-\sum_{l=0}^{d_{2}}J_{2}^{\prime}\sigma_{l}\sigma_{l+1}\\
-\sum_{m=1}^{N_{3}}J_{3}\sigma_{m}\sigma_{m+1}-\sum_{n=0}^{d_{3}}J_{3}^{\prime}\sigma_{n}\sigma_{n+1},\label{eq:H:T1D}
\end{multline}
here $J_{n}$ is the exchange interaction between atomic spins at the sites of the nearest neighbors of the original lattice in the direction $n$; also $J_{m}^{\prime}$ is an exchange interaction both between neighboring decorating spins and between neighboring decorating and nodal spins in the $m$ direction; the symbol $\sigma_{i}$ ($\sigma_{i,j}$) denotes the values $\sigma=\pm1$ at the site~$i$ ($i,j$); $N_{n}$ is the number of original lattice nodes in the direction~$n$; and $d_{m}$ is the decoration multiplicity in the direction~$m$.

Hereinafter, the Boltzmann constant ($k_{\text{B}}$) will be set equal to unity, while quantities: $T$, $J^{\prime}$ will be measured in the units of~$|J|$, as is done in the theory of low-dimensional systems.

\begin{figure*}[ht]
\centering \includegraphics{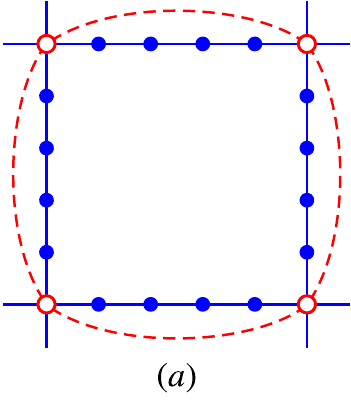}\quad{}\includegraphics{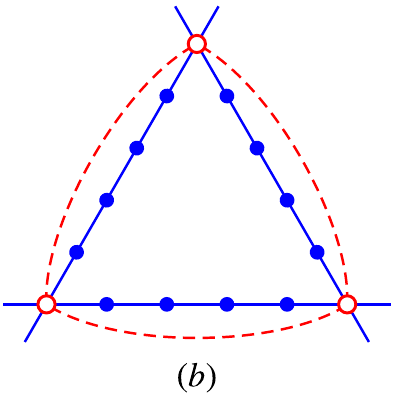}\quad{}\includegraphics{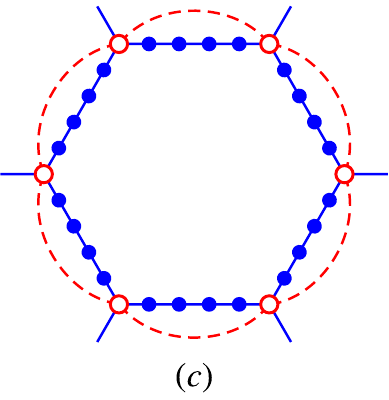}
\caption{The square (\emph{a}), triangular (\emph{b}) and honeycomb (\emph{c}) lattices, quadruply decorated in all directions. In the figure, empty red circles mark nodal spins, and filled blue circles show decorating spins. The dashed red lines indicate the interaction between spins at the nodes of the original lattice~($J$), and the solid blue lines demonstrate the interactions between decorating spins and also between decorating and nodal spins~($J^{\prime}$)}
\label{fig:TxD:lattices}
\end{figure*}

Next, we will consider only the isotropic case, 
\[
J_{n}=J,\quad d_{m}=d,\quad J_{m}^{\prime}=J^{\prime},
\]
with the same exchange interactions between spins at the sites ($J$) of the original lattice (nodal spins) in all directions, the same decoration multiplicity ($d$) in all directions, and the same exchange interactions between decoration spins and decoration and nodal spins~($J^{\prime}$).

Such an isotropic situation is shown in Fig.~\ref{fig:TxD:lattices}, where the square, triangular and honeycomb quadruply decorated lattices in all directions are demonstrated.

Using the approach of the decoration-iterative transformation proposed in~\cite{Syozi:1951}, we obtained analytical expressions for the principal (the only one maximal real) value of the Kramers--Wannier transfer-matrix~($\lambda$) for the square 
\begin{multline}
\ln\frac{\lambda_{\text{sq}}}{2}=\frac{1}{8\pi^{2}(1+2d)}\iintop_{0}^{2\pi}\ln[C^{2}\\
-DS(\cos\alpha+\cos\beta)]\,d\alpha\,d\beta,\label{eq:T2D:L}
\end{multline}
triangular
\begin{multline}
\ln\frac{\lambda_{\text{tr}}}{2}=\frac{1}{8\pi^{2}(1+3d)}\iintop_{0}^{2\pi}\ln[C^{3}+S^{3}\\
-D^{2}S(\cos\alpha+\cos\beta+\cos(\alpha+\beta))]\,d\alpha\,d\beta,\label{eq:T1D:L}
\end{multline}
and honeycomb
\begin{multline}
\ln\frac{\lambda_{\text{h}}}{2}=\frac{1}{8\pi^{2}(2+3d)}\iintop_{0}^{2\pi}\ln[2^{-1}(C^{3}+D^{3}\\
-DS^{2}(\cos\alpha+\cos\beta+\cos(\alpha+\beta)))]\,d\alpha\,d\beta\label{eq:T3D:L}
\end{multline}
decorated lattices, wherever 
\[
D=\cosh^{2+2d}\frac{J^{\prime}}{T}-\sinh^{2+2d}\frac{J^{\prime}}{T},
\]
\begin{multline*}
C=\frac{1}{2}e^{2\frac{J}{T}}\left(\cosh^{1+d}\frac{J^{\prime}}{T}+\sinh^{1+d}\frac{J^{\prime}}{T}\right)^{2}\\
+\frac{1}{2}e^{-2\frac{J}{T}}\left(\cosh^{1+d}\frac{J^{\prime}}{T}-\sinh^{1+d}\frac{J^{\prime}}{T}\right)^{2},
\end{multline*}
\begin{multline*}
S=\frac{1}{2}e^{2\frac{J}{T}}\left(\cosh^{1+d}\frac{J^{\prime}}{T}+\sinh^{1+d}\frac{J^{\prime}}{T}\right)^{2}\\
-\frac{1}{2}e^{-2\frac{J}{T}}\left(\cosh^{1+d}\frac{J^{\prime}}{T}-\sinh^{1+d}\frac{J^{\prime}}{T}\right)^{2},
\end{multline*}
and $T$ is the absolute temperature. 

Knowing the principal value of the transfer-matrix, we can calculate the specific entropy 
\begin{equation}
s=\ln\lambda+\frac{T}{\lambda}\frac{\partial\lambda}{\partial T},\label{eq:s}
\end{equation}
as well as the magnetic specific heat capacity 
\begin{equation}
c=2\frac{T}{\lambda}\frac{\partial\lambda}{\partial T}+\frac{T^{2}}{\lambda}\frac{\partial^{2}\lambda}{\partial T^{2}}-\frac{T^{2}}{\lambda^{2}}\left(\frac{\partial\lambda}{\partial T}\right)^{2}.\label{eq:c}
\end{equation}
We also generalized the spontaneous magnetization determined as the square root from pairwise spin-spin correlation function at distance between spins going to infinity
\begin{equation}
M=\sqrt{\lim_{\Delta\to\infty}\langle\sigma_{1}\sigma_{1+\Delta}\rangle},\label{eq:m}
\end{equation}
(see, for example Refs.~\cite{Montroll:1963,Baxter:2011}).

The generalized expressions for the decorated lattices (in the isotropic case) are accordingly: square
\begin{equation}
M_{\text{sq}}=\left[1-\left(\frac{D}{S}\right)^{4}\right]^{1/8},\label{eq:T2D:m}
\end{equation}
triangular 
\begin{equation}
M_{\text{tr}}=\left[1-\frac{D^{6}}{S^{3}(2C^{3}+2S^{3}+3D^{2}S)}\right]^{1/8},\label{eq:T1D:m}
\end{equation}
and honeycomb
\begin{equation}
M_{\text{h}}=\left[1-\frac{D^{3}(2C^{3}+2D^{3}+3DS^{2})}{S^{6}}\right]^{1/8}.\label{eq:T3D:m}
\end{equation}

In the particular case of zeroth decorating exchange interaction ($J^{\prime}=0$), these our formulas, are, of course, in one-to-one correspondence with the results for the square~\cite{Yang:1952}, triangular~\cite{Potts:1952}, and honeycomb~\cite{Syozi:1955} lattices, which can easily be checked by straighforward calculations.

\section{Types of frustrated systems}

It should be clarified that the partition function of the Ising model on any 2D lattice is a sum of $2^N$ configurations with all possibly fixed spin values ($+1$ or $-1$) with Boltzmann statistical weights. At infinite temperature, the weights of all configurations are equal, which means complete degeneracy and the entropy per spin~(\ref{eq:s}) is equal to
\[
\lim_{T\to\infty } s \equiv s^{\infty } = \ln 2 \approx 0.693\,15.
\] 
Obviously, the sublogarithmic expression is equal to the statistical weight ($W=2$) determined by the number of states on the node in the considered Ising model.

With temperature lowering, the weights of some configurations increase (survive), the others decrease, and the degeneracy lowers and lowers down to zero temperature, accompanied by continuous decrease of entropy. Depending on concrete set of exchange interaction parameters between atomic spins and the multiplicity of decoration of the considered planar lattices, three possible scenarios for the existence of various magnetic states may appear in the system, which are determined by the features of frustrations and the ordering of the spin system. 

Next, for clarity, we will discuss the results for the following set of model parameters, in which the exchange interactions between nodal and decorating spins are equal to $|J|=|J^{\prime}|=1$.


\emph{First scenario}.
At zero temperature, only one configuration survives, the degeneracy disappears, and a magnetically ordered equilibrium state arises in the system.
In this case, the zero-temperature (residual) entropy is equal to zero
\begin{equation}
s^{\circ}=\ln 1=0,
\label{eq:s:eq:0}
\end{equation}
the statistical weight of this state is equal to one ($W=1$), and the system is non-degenerate, which corresponds to the Nernst--Planck theorem~\cite{Sommerfeld:1956,Nolting8:2018}.

In this scenario, the survived configuration possesses translational invariance, which means a complete long-range magnetic order at $T=0$.
The system experiences a magnetic phase transition at a non-zero temperature ($T_{\text{c}}>0$).

The heat capacity undergoes logarithmic divergence ($\lambda$-shaped peak) at the magnetic phase transition point ($T_{\text{c}}$), that corresponds to the appearance of long-range magnetic order.
The temperature dependence of the heat capacity can have both a single-peak and a two-peak structure, when the second peak has a dome-shaped (see the discussion of this effect in Refs.~\cite{Zarubin:2020,Kassan-Ogly:2019,Proshkin:2019,Kassan-Ogly:2018:}).

In this case, the zero-temperature (residual) spontaneous magnetization of the system is maximum (saturated) and equals unity, $M^{\circ}=1$.
As the temperature rises, the spontaneous magnetization (\ref{eq:m}) disappears at the phase transition point ($T_{\text{c}}$).
It should be noted that such a behavior is intrinsic to all undecorated lattices.

This type of behavior of decorated spin systems is observed at ferro-/ferromagnetic types of exchange interactions $(J=+1$, $J^{\prime}=+1$) in the system and any decoration multiplicity ($d>0$) of square, triangular, and honeycomb lattices; at ferro-/antiferromagnetic types of exchange interactions ($J=+1$, $J^{\prime}=-1$) and odd values of decoration multiplicity ($d>0$) for all three types of decorated planar lattices; also at antiferro-/antiferromagnetic types of exchange interactions ($J=-1$, $J^{\prime}=-1$) and even values of decoration multiplicity ($d>1$) for square and honeycomb lattices. 

\begin{figure}[ht]
\centering \includegraphics{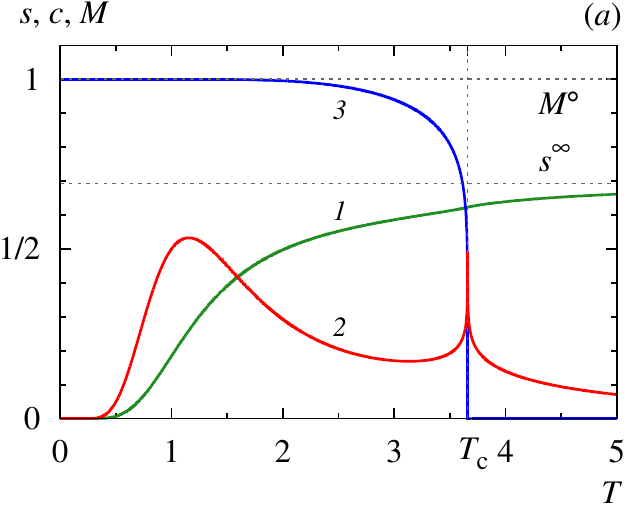}\quad{}\includegraphics{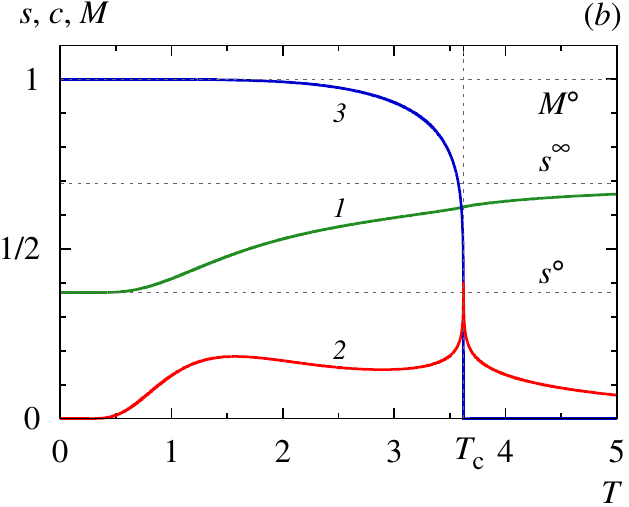}\quad{}\includegraphics{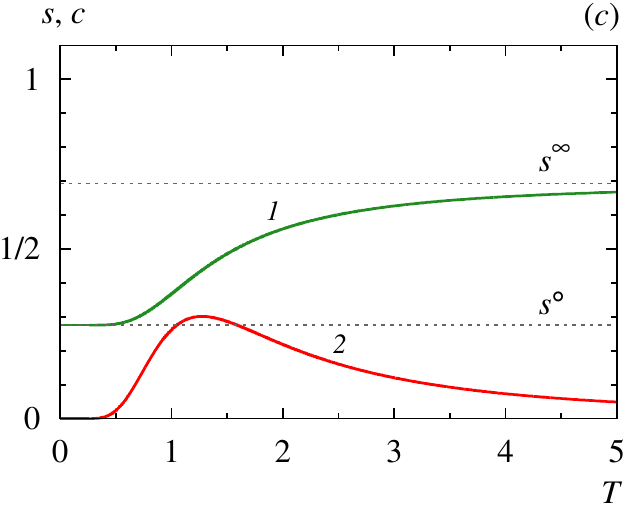}
\caption{Specific entropy (green line~\emph{1}), heat capacity (red line~\emph{2}), and spontaneous magnetization (blue line~\emph{3}) for quadruply decorated ($d=4$) triangular lattice at ferro-/ferromagnetic exchange interactions ($J=+1$, $J^{\prime}=+1$), where $s^{\circ}=0$, $M^{\circ}=1$, $T_{\text{c}}\approx 3.658\,93$~(\emph{a}); at ferro-/antiferromagnetic exchange interactions ($J=+1$, $J^{\prime}=-1$), where $s^{\circ}\approx 0.371\,41$, $M^{\circ}\approx 0.999\,87$, $T_{\text{c}}\approx 3.622\,28$~(\emph{b}); and at antiferro-/ferromagnetic exchange interactions ($J=-1$, $J^{\prime}=+1$), where $s^{\circ}\approx 0.275\,54$~(\emph{c})}
\label{fig:T1D:TD}
\end{figure}

\begin{figure}[ht]
\centering \includegraphics{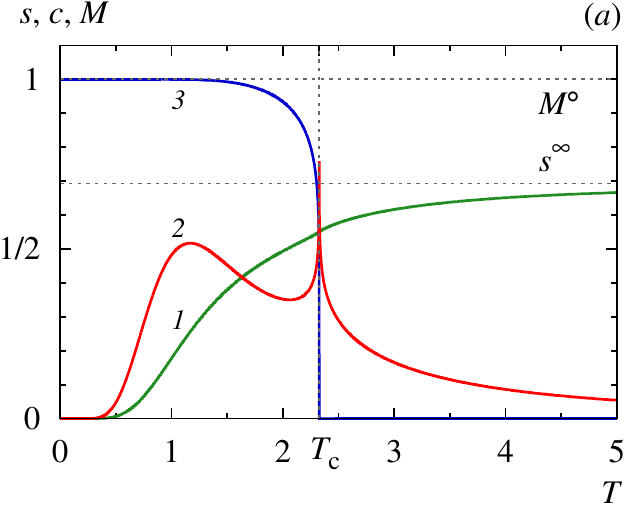}\quad{}\includegraphics{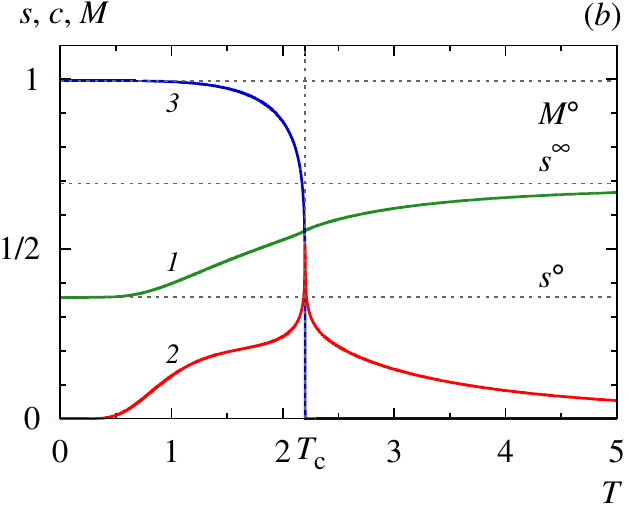}
\caption{Specific entropy (green line~\emph{1}), heat capacity (red line~\emph{2}), and spontaneous magnetization (blue line \emph{3}) for quadruply decorated ($d=4$) square lattice at ferro-/ferromagnetic exchange interactions ($J=+1$, $J^{\prime}=+1$), where $s^{\circ}=0$, $M^{\circ}=1$, $T_{\text{c}}\approx 2.326\,81$~(\emph{a}); and at ferro-/antiferromagnetic exchange interactions ($J=+1$, $J^{\prime}=-1$), where $s^{\circ}\approx 0.357\,85$, $M^{\circ}\approx 0.996\,18$, $T_{\text{c}}\approx 2.199\,38$~(\emph{b})}
\label{fig:T2D:TD}
\end{figure}

\begin{figure}[ht]
\centering \includegraphics{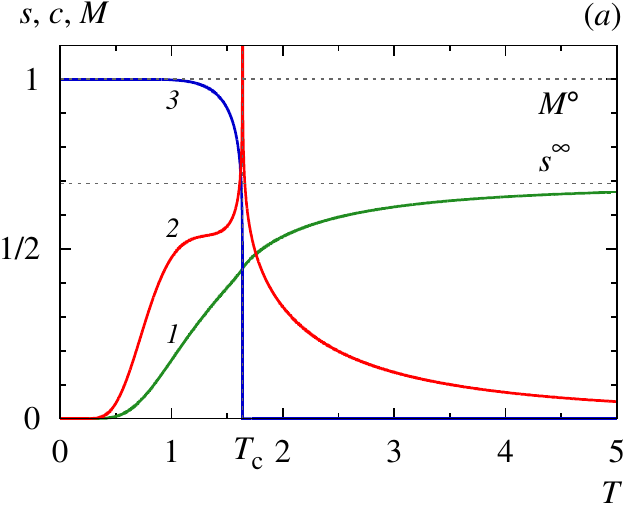}\quad{}\includegraphics{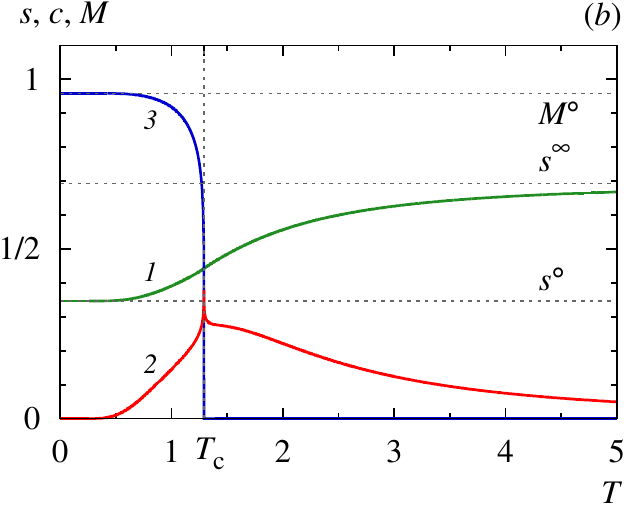}
\caption{Specific entropy (green line~\emph{1}), heat capacity (red line~\emph{2}), and spontaneous magnetization (blue line \emph{3}) for quadruply decorated ($d=4$) honeycomb lattice at ferro-/ferromagnetic exchange interactions ($J=+1$, $J^{\prime}=+1$), where $s^{\circ}=0$, $M^{\circ}=1$, $T_{\text{c}}\approx 1.637\,81$~(\emph{a}); and at ferro-/antiferromagnetic exchange interactions ($J=+1$, $J^{\prime}=-1$), where $s^{\circ}\approx 0.346\,60$, $M^{\circ}\approx 0.958\,96$, $T_{\text{c}}\approx 1.290\,75$~(\emph{b})}
\label{fig:T3D:TD}
\end{figure}

This scenario is shown in Figs.~\ref{fig:T1D:TD}\emph{a}, \ref{fig:T2D:TD}\emph{a}, \ref{fig:T3D:TD}\emph{a}, where plots of the behavior of thermodynamic functions in this mode for the corresponding decorated planar lattices are presented.


Further, it should be noted that, depending on the decoration multiplicity of the lattices under consideration and the types of exchange interactions between nodal and decorating spins, competing exchange interactions may arise in the system.

It is these competing exchange interactions in spin systems that form the regimes of magnetic frustrations. The presence of such frustrations is accompanied by a nonzero residual entropy of the system 
\begin{equation}
s^{\circ}=\ln W>0.
\label{eq:s:g:0}
\end{equation}
That means that the statistical weight of this state is greater than unity ($W>1$), and the system itself is degenerate~\cite{Nolting8:2018}.

In this model, the state of the system, in which the entropy of the ground state is greater than zero, $0<s^{\circ}\leqslant s^{\infty}$, while the statistical weight of the system is in the range $1<W\leqslant 2$, should be referred to as \emph{frustrated}. 
Note that this situation does not contradict the third law of thermodynamics~\cite{Sommerfeld:1956} and was thoroughly discussed in Refs.~\cite{Zarubin:2019:,Zarubin:2019:E}.

As an example, we present the dependence of the residual entropy on the decoration multiplicity for an isotropic decorated square lattice
\begin{equation}
s^{\circ}=\frac{1}{2\pi(1+2d)}\intop_{0}^{\pi}\ln\frac{\tau}{2}\,d\alpha,
\label{eq:T2D:s:0}
\end{equation}
\begin{multline*}
\tau=(d^{2}+2d+2)^{2}\\
+\sqrt{(d^{2}+2d+2)^{4}-16d^{2}(d+1)^{2}(d+2)^{2}\cos^{2}\alpha}.
\end{multline*}
Several cases where this formula is applicable are described below.

Depending on whether competing exchange interactions lead to complete magnetic disorder or to partial magnetic ordering of the spin system, the model makes it possible to describe different types of behavior of the heat capacity and spontaneous magnetization.


\emph{Second scenario}. 
At zero temperature, the competing exchange interactions lead to complete magnetic disorder, the infinite number of configurations with equal statistical weights, but without translational invariance, survive, the degeneracy does not disappear, and the residual entropy is nonzero~(\ref{eq:s:g:0}).

The system experiences frustrations that destroy the long-range magnetic order, as a result of which the system does not experience a magnetic phase transition.
The heat capacity temperature dependence has one dome-shaped peak and does not undergo divergence.
There is no spontaneous magnetization of the system, and the the phase transition is absent.
Such a behavior is intrinsic also to undecorated lattices. 

Full antiferromagnetic Ising model on triangular and kagome lattices was discussed in original papers by Wannier~\cite{Wannier:1950} and Kano and Naya~\cite{Kano:1953}.

This type of behavior of decorated spin systems is observed at antiferro-/antiferromagnetic types of exchange interactions ($J=-1$, $J^{\prime}=-1$) in the system and any multiplicity of decoration ($d>0$) of the triangular lattice and the only case when $d=1$ for square and honeycomb lattices; at antiferro-/ferromagnetic types of exchange interactions ($J=-1$, $J^{\prime}=+1$) and any multiplicity of decoration ($d>0$) of a triangular lattice and a single case with $d=1$ for a decorated square lattice and two cases with $d=1$ and $d=2$ for a decorated honeycomb lattice; also at ferro-/antiferromagnetic types of exchange interactions ($J=+1$, $J^{\prime}=-1$) and only one decoration multiplicity at $d=2$ for a honeycomb lattice.

In Fig.~\ref{fig:T1D:TD}\emph{c} the behavior of thermodynamic functions in this mode for a decorated triangular lattice are shown.


\emph{Third scenario}.
In a number of cases, a competing exchange interactions can lead to partial magnetic ordering, while the infinite number of configurations are survived at zero temperature, the degeneracy does not disappear, and the residual entropy is nonzero~(\ref{eq:s:g:0}).
Interestingly, in the case of partial magnetic ordering, the frustrated system experiences a phase transition.

The heat capacity of the system diverges at the phase transition point, i.e. it has a $\lambda$-shaped peak at the critical temperature of the magnetic phase transition ($T_{\text{c}}>0$), which corresponds to the appearance of long-range magnetic order.
As in the first scenario, the temperature dependence of the heat capacity can have both a single-peak and a two-peak structure, when the second peak is dome-shaped.

Spontaneous magnetization with increasing temperature vanishes at the transition point ($T_{\text{c}}$). Moreover, it is important to note that due to the partial magnetic ordering of the spin system, the residual spontaneous magnetization does not reach saturation at zero temperature and is less than unity, $M^{\circ}<1$. 

Let us also give an example of the dependence of the residual spontaneous magnetization on the decoration multiplicity for an isotropic decorated square lattice 
\begin{equation}
M^{\circ}=\left[1-\left(\frac{2(d+1)}{d(d+2)}\right)^{4}\right]^{1/8}.\label{eq:T2D:M:0}
\end{equation}

\begin{figure}[ht]
\centering \includegraphics{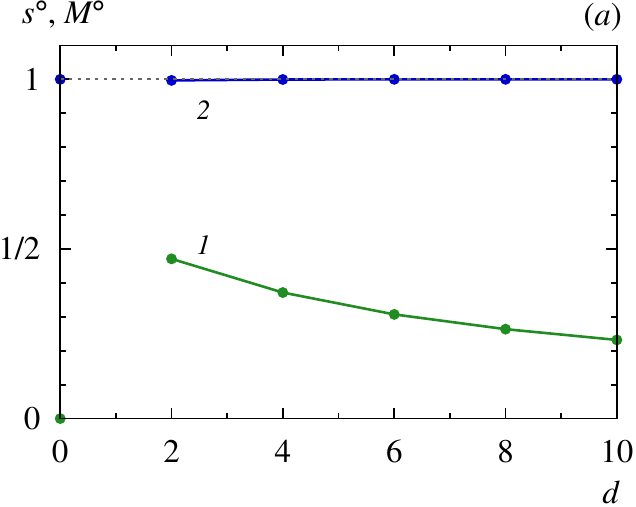}\quad{}\includegraphics{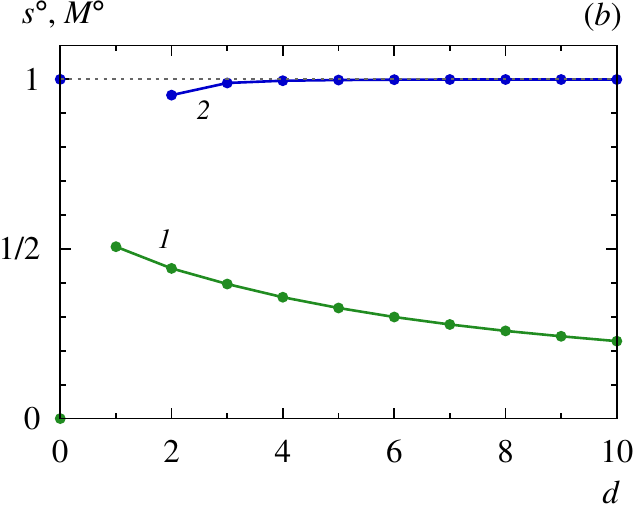}\quad{}\includegraphics{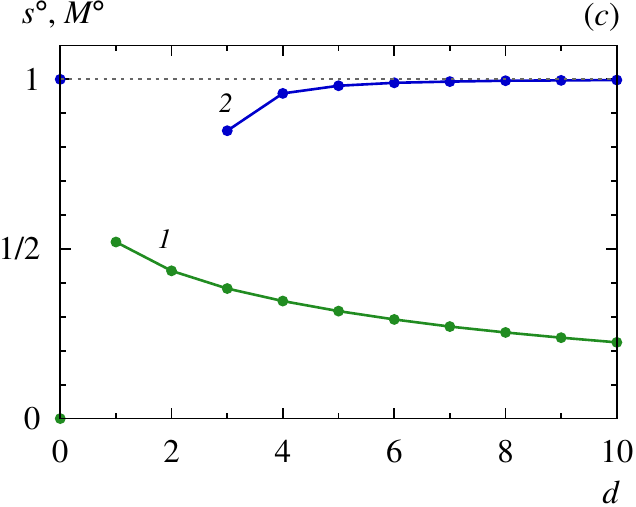}
\caption{Dependence of the residual entropy (green line~\emph{1}) and residual spontaneous magnetization (blue line~\emph{2}) on the decoration multiplicity for a decorated triangular lattice (\emph{a}) at ferro-/antiferromagnetic exchange interactions ($J=+1$, $J^{\prime}=-1$), also for decorated triangular (\emph{b}) and honeycomb (\emph{c}) lattices at antiferro-/ferromagnetic exchange interactions ($J=-1$, $J^{\prime}=+1$)}
\label{fig:TxD:TD:0}
\end{figure}

With an increase in the decoration multiplicity, the value of residual entropy asymptotically approaches zero and residual spontaneous magnetization asymptotically approaches unity,
\[
\lim_{d\to\infty}s^{\circ}\to0,\quad\lim_{d\to\infty}M^{\circ}\to1.
\]
Let us give examples of residual entropy and residual spontaneous magnetization at different values of the degree of decoration of planar lattices, for example, at antiferro-/ferromagnetic exchange parameters ($J=-1$, $J^{\prime}=+1$) for a decorated square lattice 
\[
s^{\circ}(d=2)\approx0.442\,66,\quad s^{\circ}(d=10)\approx0.228\,37,
\]
\[
M^{\circ}(d=2)\approx0.953\,56,\quad M^{\circ}(d=10)\approx0.999\,86,
\]
at ferro-/antiferromagnetic exchange parameters ($J=+1$, $J^{\prime}=-1$) for decorated triangular lattice 
\[
s^{\circ}(d=2)\approx0.471\,04,\quad s^{\circ}(d=6)\approx0.307\,25,
\]
\[
M^{\circ}(d=2)\approx0.997\,04,\quad M^{\circ}(d=6)\approx0.999\,98,
\]
and at antiferro-/ferromagnetic exchange parameters ($J=-1$, $J^{\prime}=+1$) for decorated honeycomb lattice 
\[
s^{\circ}(d=3)\approx0.383\,14,\quad s^{\circ}(d=10)\approx0.224\,86,
\]
\[
M^{\circ}(d=3)\approx0.848\,86,\quad M^{\circ}(d=10)\approx0.997\,93.
\]
In Fig.~\ref{fig:TxD:TD:0} the dependence of the residual entropy and residual spontaneous magnetization on the lattice decoration multiplicity for the studied types of decorated planar lattices are shown.

It should be noted that in the described case, both magnetic frustrations and a magnetic phase transition coexist in the presented decorated planar lattices. It is usually assumed that in frustrated systems, due to competing exchange interactions, the emerging magnetic frustrations suppress the long-range magnetic order, as a result of which no phase transition is observed in the system as the temperature changes. In this regime, it can be seen that such coexistence is possible, since the spin system has a partial magnetic ordering, which is accompanied by an unsaturated residual spontaneous magnetization. 
Such behavior is unusual for undecorated lattices.

This type of behavior of decorated spin systems is observed at antiferro-/ferromagnetic types of exchange interactions ($J=-1$, $J^{\prime}=+1$) in the system and any decoration multiplicity square (for $d>1$) and honeycomb (for $d>2$) lattices; at antiferro-/antiferromagnetic types of exchange interactions ($J=-1$, $J^{\prime}=+1$) and any decoration multiplicity ($d>1$) of a square lattice and odd values of decoration multiplicity ($d>2$) of a honeycomb lattice; also at ferro-/antiferromagnetic types of exchange interactions ($J=+1$, $J^{\prime}=-1$) and even values of decoration multiplicity for square, triangular and honeycomb (for $d>2$) lattices. 

In Figs.~\ref{fig:T1D:TD}\emph{b}, \ref{fig:T2D:TD}\emph{b}, \ref{fig:T3D:TD}\emph{b} the plots of the behavior of thermodynamic functions in this mode for the corresponding decorated planar lattices are shown. 


\begin{figure}[ht]
\centering \includegraphics{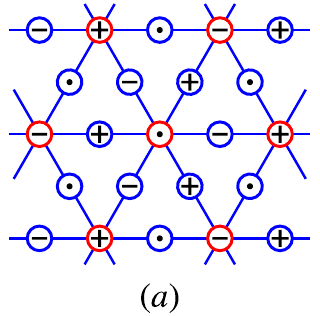}\quad \includegraphics{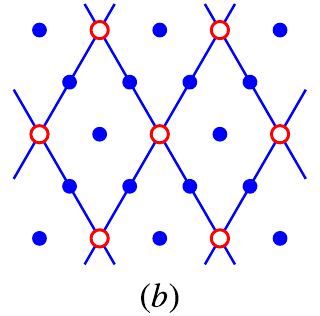}
\caption{Triangular lattice, singly decorated in all three directions with one of the possible spin configurations (\emph{a}), where the plus and minus signs, respectively, mark positive and negative spins, and the dot shows the positions of frustrated states in which each sign individually may be picked random; and with one (horizontal) exchange interaction is equal to zero (\emph{b})}
\label{fig:T1D:PMO}
\end{figure}

Let us give two examples of partial magnetic ordering in the Ising model on a triangular lattice singly decorated in all three directions (Fig.~\ref{fig:T1D:PMO}).

In Fig.~\ref{fig:T1D:PMO}\emph{a} the instantaneous configuration of a regular sequence of spins on a singly decorated triangular lattice in all three directions is shown. In this case, it is easy to see that such a configuration of the spin system at ferro-/antiferromagnetic exchange interactions will experience frustrations that will destroy the initial spin ordering.

In Fig~\ref{fig:T1D:PMO}\emph{b} an example of another type of partial ordering in the Ising model on a triangular lattice singly decorated in all three directions is shown.
In this case, one (horizontal) exchange interaction is equal to zero. Here partial ordering is provided only by spins connected by antiferromagnetic interactions ($T_{\text{c}}\approx 1.308$). Free spins provide the degeneracy of the ground state such that residual entropy has non-zero value ($s^{\circ }\approx 0.173\,28$).

\section{Conclusions}

The paper considers the frustration properties of the Ising model on three decorated planar lattices, such as square, triangular and honeycomb. Analytical expressions are obtained for the entropy, heat capacity, and spontaneous magnetization of such systems. The effect of magnetic frustrations on the behavior of the thermodynamic functions of decorated planar spin systems is also studied.

It is shown that the geometry of decorated planar lattices, as well as the types of exchange interactions of spins and their relationship at nodal and decorated lattice sites, lead to interesting results. Even despite substantial simplifications, when the multiplicity of lattice decoration and the magnitude of exchange interactions (both between nodal and decorating spins) are the same in all directions of the lattice and are equal in absolute value to unity, spin systems experience several regimes in which the frustration phenomenon clearly manifests itself.

Situations are demonstrated when, for certain ratios of model parameters and decorated planar lattices, there are no magnetic frustrations, the residual (zero-temperature) entropy is zero, and the spin system experiences a magnetic phase transition at the point at which the heat capacity has a $\lambda$-shaped (logarithmically divergent) peak. In turn, the residual (zero-temperature) spontaneous magnetization is saturated and equal to unity, while the spontaneous magnetization vanishes at the point of the magnetic phase transition.

In the second case, frustrations completely suppress magnetic ordering and the system does not experience magnetic phase transitions, while the residual entropy is nonzero, the temperature dependence of the heat capacity has a dome-shaped peak, and there is no spontaneous magnetization.

In the third situation, the decorated spin system is frustrated, the residual entropy is not equal to zero, but the spin system has partial magnetic ordering, which leads to a phase transition at the point where the temperature dependence of the heat capacity has a $\lambda$-shaped peak. The residual spontaneous magnetization does not reach saturation and is less than unity; in this case, the spontaneous magnetization vanishes at the point of the magnetic phase transition.

Infinite sequences of residual entropies in single-integral form and infinite series of residual spontaneous magnetization in simple numbers are the crowning glory of our results.

In conclusion, we emphasize that contrary to the widespread opinion that frustration (non-zero residual entropy) and phase transition (logarithmic divergence of heat capacity) are inevitably mutually-exclusive features, we have found that although it is sometimes true, but they can well coexist. The most striking result of our study~-- in all cases, when the frustrations and phase transition are observed, they are accompanied by an unsaturated residual spontaneous magnetization that is without fail less than unity at zero temperature.

\begin{acknowledgments}
The research was carried out within the state assignment of Ministry of Science and Higher Education of the Russian Federation (theme ``Quantum'' No.~122021000038-7).
\end{acknowledgments}

\bibliographystyle{apsrev4-2}
\bibliography{residual_arxiv}

\end{document}